\documentclass[superscriptaddress,showkeys,showpacs]{revtex4}
\usepackage{graphicx}
\usepackage[tbtags]{amsmath}
\begin{document}
\title{Sea quark polarization and semi-inclusive DIS data.}

\author{X. Jiang} 
\email{jiang@jlab.org}
\affiliation{Department of Physics and Astronomy, Rutgers, the State
University of New Jersey\\ 136 Frelinghuysen Road, Piscataway, NJ 08854 USA.
}
\author{G. A. Navarro} 
\email{gabin@df.uba.ar}
\affiliation{Departamento de F\'{\i}sica,
Universidad de Buenos Aires\\ Ciudad Universitaria, Pab.1 (1428)
Buenos Aires, Argentina}
\author{R. Sassot}
\email{sassot@df.uba.ar}
\affiliation{Departamento de F\'{\i}sica,
Universidad de Buenos Aires\\ Ciudad Universitaria, Pab.1 (1428)
Buenos Aires, Argentina}
\begin{abstract}
We investigate the potential impact of forthcoming Jefferson Lab  semi-inclusive 
polarized deep inelastic scattering proton measurements in the determination 
of the sea quark polarization in the nucleon by means of a next to leading 
order global QCD analysis. Specifically, we estimate the resulting improvement 
in the constraints on polarized parton densities for the different flavors, 
which is found to be significant for up and strange quarks, and the 
correlation between remaining uncertainty ranges for each of the parton 
species.

\end{abstract}

\pacs{12.38.Bx, 13.85.Ni}
\keywords{Semi-Inclusive DIS; perturbative QCD.}

\maketitle

\section{Introduction}

The way in which sea quarks are polarized inside nucleons has been
a persisting question ever since the spin structure of the proton
began to be unveiled by polarized DIS experiments and even today, in spite of 
several successful experimental programs, remains to a large extent unanswered.
Contrary to the common belief before the paradigmatic EMC experiment at the 
end of 1980s \cite{Ashman:1987hv,Ashman:1989ig}, the data obtained
by the collaboration suggested that sea quarks and gluons in the nucleon 
carried non negligible polarization. However this conclusion was, and has 
been for many years, conditional upon rather strong assumptions on isospin 
symmetry extended to polarized phenomena. In the subsequent years, isospin 
symmetry itself become seriously questioned \cite{Kumano:1997cy} and 
consequently the sea quark polarization turned into an even more elusive 
question.

Global QCD analyses including semi-inclusive measurements of polarized 
lepton-nucleon deep inelastic processes began to change this situation 
more recently \cite{deFlorian:1997ie,deFlorian:2000bm}, and today these 
data allow to constrain the extraction of polarized parton densities in 
QCD global fits \cite{deFlorian:2005mw}. The effectiveness of these 
constraints of course relies on the precision of the data and this is why 
the forthcoming generation of semi-inclusive experiments is crucial.

In a QCD global fit the uncertainty range of the resulting parton densities 
can be estimated by analyzing the profile of the $\chi^2$-function of the fit 
to data against variations in the different features of the densities. This 
technique has been widely used in extractions of unpolarized parton densities 
\cite{martin:2003} and more recently has been implemented in the polarized case \cite{deFlorian:2005mw} providing reliable constraints on their different 
features, such as the net polarization carried by each parton flavor. 

Of course, the values each flavor polarization, or some other parameter, may 
take within these constraints are not independent, but become correlated.
Even in the case where there are enough independent observables to extract
in principle all the parton densities, the uncertainties in the measurements
of those observables, together with the theoretical uncertainties inherent 
in the fitting procedure, conspire against the independence of parton densities
and results in correlations. A strong correlation between two parton densities,
consequently means that neither of them is actually well determined. The 
inclusion of new and more precise data should not only reduce the uncertainty
ranges for each flavor but also those correlations.  

In this article we investigate the impact of the inclusion of a
series of semi-inclusive deep inelastic scattering measurements 
to be performed at Jefferson Lab  \cite{xiaodong} in a next to leading order 
(NLO) QCD global fit to all the available inclusive and semi-inclusive deep 
inelastic scattering data. In order to do this we take into account
the kinematical coverage, statistics and level of uncertainty
expected for the measurements. We also analyze the correlation
between the uncertainty ranges of the different sea quark
polarizations. As result of this analysis we find that the forthcoming 
Jefferson Lab  experiment will effectively contribute to constrain the 
sea quark
polarization in the proton. The most significant improvement is found
in up sea quark distributions, and also with a noticeable effect for
strange quarks. The improvement in both distributions will be
related to the inclusion in the fit of more precise data on charged meson
electroproduction. Another interesting effect of the inclusion of
these data is the softening of the correlation between the allowed range for
the polarization of up and down sea quarks.

In the following section, we settle definitions and conventions
for the global fitting procedure and the way in which we study the
profile of the $\chi^2$ function, we discuss the characteristics
of the forthcoming semi-inclusive experiment, and explain how we evaluate
the impact of it in a global fit. Then, we compare the results 
coming from the analysis of the set of data available at present, against 
those that would come from the data set enlarged with the forthcoming 
measurements, both for the individual uncertainty 
ranges, for the net polarization of the different flavors, and for the
correlations between flavors. Finally, we present our conclusions.

\section{Global QCD fits and new data}

In the present analysis we implement the NLO QCD global fit to existing 
data along the lines of what was done in reference 
\cite{deFlorian:2005mw} but restricting the input fragmentation functions to 
those of reference \cite{kretzer}, which were shown to give the best fits 
to combined polarized data. The NLO expressions for both inclusive and 
semi-inclusive spin-dependent asymmetries and evolution equations for the 
parton densities can be found in  \cite{inclusiva,newgr} and \cite{NPB} 
respectively. 

The data sets analyzed  include only points with $Q^2>1$ GeV$^2$,
listed in Table \ref{tab:table1},  and totaling 137, 139, and 37 points, from 
proton, deuteron, and helium targets respectively, from polarized  inclusive
deep inelastic scattering  plus 60, 87, and 
18, from proton, deuteron, and helium targets respectively from semi-inclusive
deep inelastic scattering.

The main conclusions reached in reference 
\cite{deFlorian:2005mw} were that using the Lagrange multiplier approach 
\cite{Stump:2001gu} as a mean to explore the 
profile of the $\chi ^2$ function against different degrees of polarization 
in each parton flavor, definite estimates for the uncertainty in the net 
polarization 
of each flavor, and in the parameters of the polarized parton distributions, 
can be obtained. 
The overall result is a well constrained scenario where semi-inclusive data 
is not only consistent with inclusive measurements, but improves the 
constraining power of the fit for all the distributions, being crucial for 
the light sea quarks. 

The best fits suggest an overall picture for the quark 
densities at  $10$ GeV$^2$ where, within uncertainties, up quarks are almost 
$100\%$ polarized parallel to the proton, down quarks anti-parallel in a 
similar proportion, and sea quarks have a small and  flavor symmetric negative 
polarization. The first moment of the gluon distribution is found to be in 
agreement with the most recent direct measurements \cite{Ellis:2005cy}
close to 0.6, constrained to be smaller than 0.8 and larger than -0.05 
within a conservative increase in the $\chi ^2$ value within a two percent 
range ($\Delta \chi^2=2\%$). 

\begin{table}
\caption{\label{tab:table1} Inclusive and semi-inclusive data used in the fit.}
\begin{ruledtabular}
\begin{tabular}{ccccc} 
Collaboration & Target& Final state & \# points & Refs. \\ \hline
EMC          & proton& inclusive   &    10     & \cite{Ashman:1989ig} \\ 
SMC          & proton, deuteron & inclusive &  12, 12  & \cite{SMCi} \\  
E-143        & proton, deuteron & inclusive  &  82, 82  & \cite{E143} \\ 
E-155        & proton, deuteron & inclusive   &    24, 24    & \cite{E155} \\ 
Hermes       & proton,deuteron,helium& inclusive   &    9, 9, 9   & \cite{HERMES} \\
E-142        & helium& inclusive   &    8     & \cite{E142} \\ 
E-154        & helium& inclusive   &    17     & \cite{E143} \\ 
Hall A       & helium & inclusive &   3      & \cite{HALLA}   \\
COMPASS      & deuteron & inclusive &   12  & \cite{COMPASS} \\
 \hline 
SMC          & proton,deuteron& $h^+$, $h^-$  &  24, 24  & \cite{SMC} \\ 
Hermes       & proton, deuteron, helium & $h^+$, $h^-$, $\pi^+$, $\pi^-$, $K^+$, $K^-$, $K^T$   &    36,63,18     
& \cite{HERMES} \\  \hline
\multicolumn{3}{c}{Total} & 478 & \\
\end{tabular}
\end{ruledtabular}
\end{table}

In order to evaluate the impact of the forthcoming semi-inclusive proton data of 
Jefferson Lab, we included in the global analysis the expected values of the $\pi^+$, $\pi^-$, $K^+$ and $K^-$ semi-inclusive asymmetries on a polarized proton target, 
computed with the 
best set of parton densities obtained in \cite{deFlorian:2005mw} with expected 
experimental uncertainties, as an additional set of points to be fitted. 
The projected statistical accuracies of these asymmetries are based on a total of 225 hours of 6 GeV polarized electron beam on a polarized $NH_3$ target. The electron beam current is assumed to be 80 nA with a polarization of 80 \%. The standard Jefferson Lab Hall C polarized $NH_3$ target of 3 cm thickness and 80 \% polarization is assumed. The scattered electron will be detected at 30 degree with an array of lead-glass detectors in conjunction with a threshold gas Cherenkov counter, covering a solid angle of 210 msr. The produced hadron will be detected in coincidence using the standard Hall C High Momentum Spectrometer (HMS) at 10.8 degree and a central momentum of 2.7 GeV/c ($z_{\pi} \approx 0.5$). The HMS spectrometer has a solid angle of 6 msr and a momentum acceptance of $\pm 10 \%$.

With this enlarged set of asymmetries to be fitted, we have redone the 
analysis of \cite{deFlorian:2005mw} adding a detailed study of the 
correlations, and compared the resulting constraints on 
polarization with those of the original fit. 

\section{Results}

\setlength{\unitlength}{1.mm}
\begin{figure}[b]
\includegraphics[width=12cm]{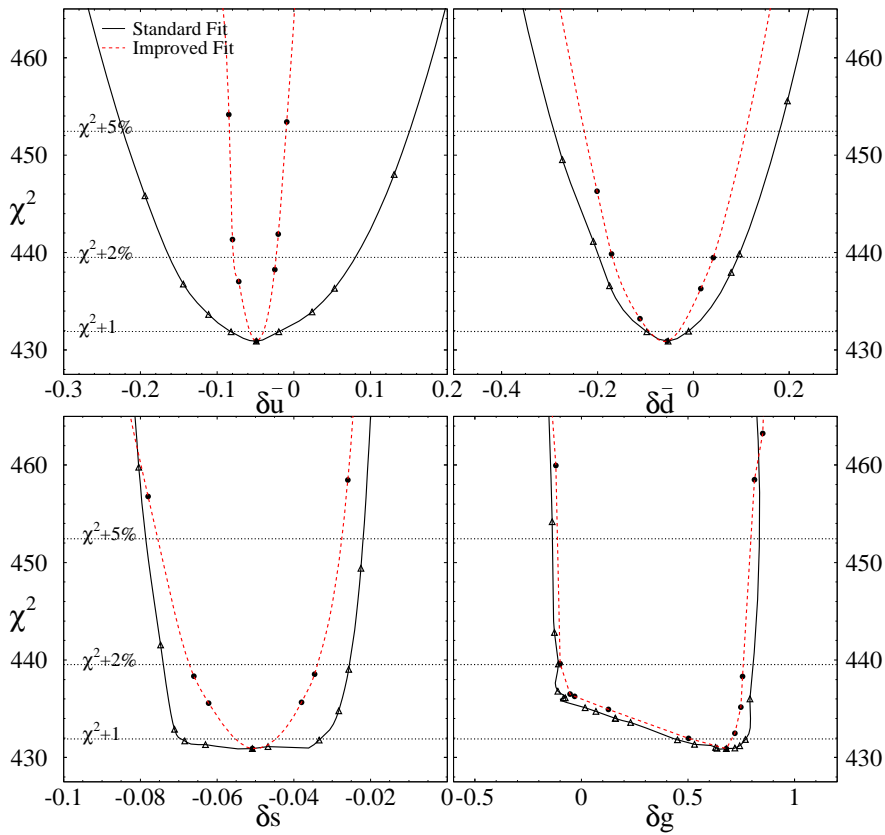}
\caption{Profile of the $\chi^2$ function against parton polarizations.}
\label{fig:par}
\end{figure}

We start with the estimates for the uncertainties in the polarization of the 
different quark flavors. In Figure \ref{fig:par} we show the outcome of varying
the $\chi^2$ of the NLO fits with the set of data available at present, in the 
following referred to as ``standard fit'', and the ``improved fit'', which 
includes the asymmetries expected to be measured by the E04-113, against the 
first moment of the respective polarized parton densities 
$\delta \overline{q}$ at $Q^2=10$ GeV$^2$, one at a time. This is, to minimize
\begin{equation}
\Phi(\lambda_q, a_j)=\chi^2(a_j)+\lambda_q\, \delta q(a_j) \,\,\,\,\,\,\,\,\,\,\,\, q=u,\overline{u},d,\overline{d},s,g.
\end{equation}
where $\lambda_q$ is the Lagrange multiplier associated to the
polarization of a given quark flavor $q$, $a_j$ are the parameters to be 
fitted, and the $\chi^2$ definition is the most simple and commonly used in 
fits to polarized data,  namely, 
\begin{equation}
\chi^2=\sum_{i=1}^N \frac{(T_i(a_j)-E_i)^2}{\sigma_i^2}\,.
\end{equation}
\setlength{\unitlength}{1.mm}
\begin{figure}[hbt]
\includegraphics[width=14cm]{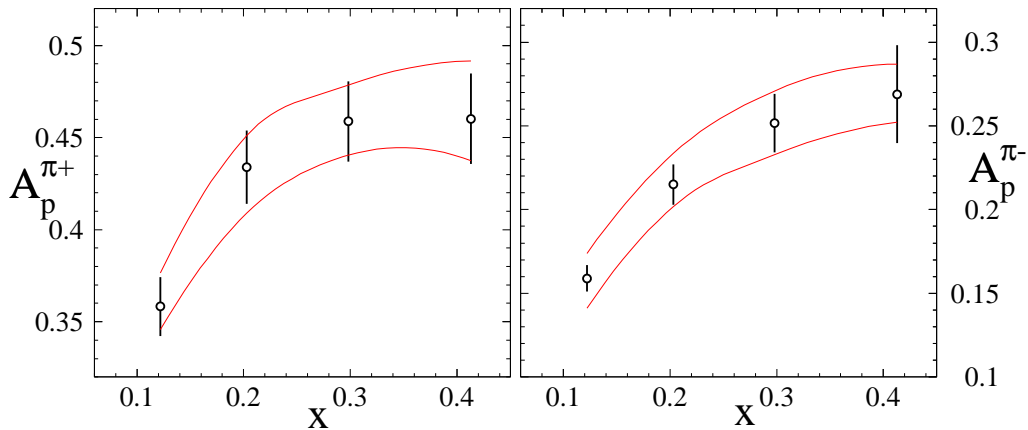}
\caption{Uncertainty bands for  $A_p^{\pi +}$ and  $A_p^{\pi -}$  asymmetries
coming the ``improved fit'' and E04-113 expected uncertainties.}
\label{fig:new}
\end{figure}
In Equation (2), $E_i$ is the measured value of a given observable, 
$T_i$ is the
corresponding theoretical estimate computed with a given set of parameters
for the polarized parton densities, and $\sigma_i$ is the error associated 
with the measurement, usually taken to be the addition of the reported 
statistical and systematic errors in quadrature. Notice that the additional 
set of asymmetries included does not contribute to $\chi^2$ when it is computed with 
parton densities corresponding to the best fit of reference 
\cite{deFlorian:2005mw}, the same densities used to generate the asymmetries, 
situation that happens at the minima of the curves. As the distributions 
change in 
order to increase or reduce the polarization of a given flavor, the $\chi^2$
obtained with one or another set begin to differ. The solid lines in 
Figure \ref{fig:par} correspond to the analysis of the standard set of data,
while the dashed lines includes the estimated impact of future measurements.
\setlength{\unitlength}{1.mm}
\begin{figure}[hbt]
\includegraphics[width=19cm]{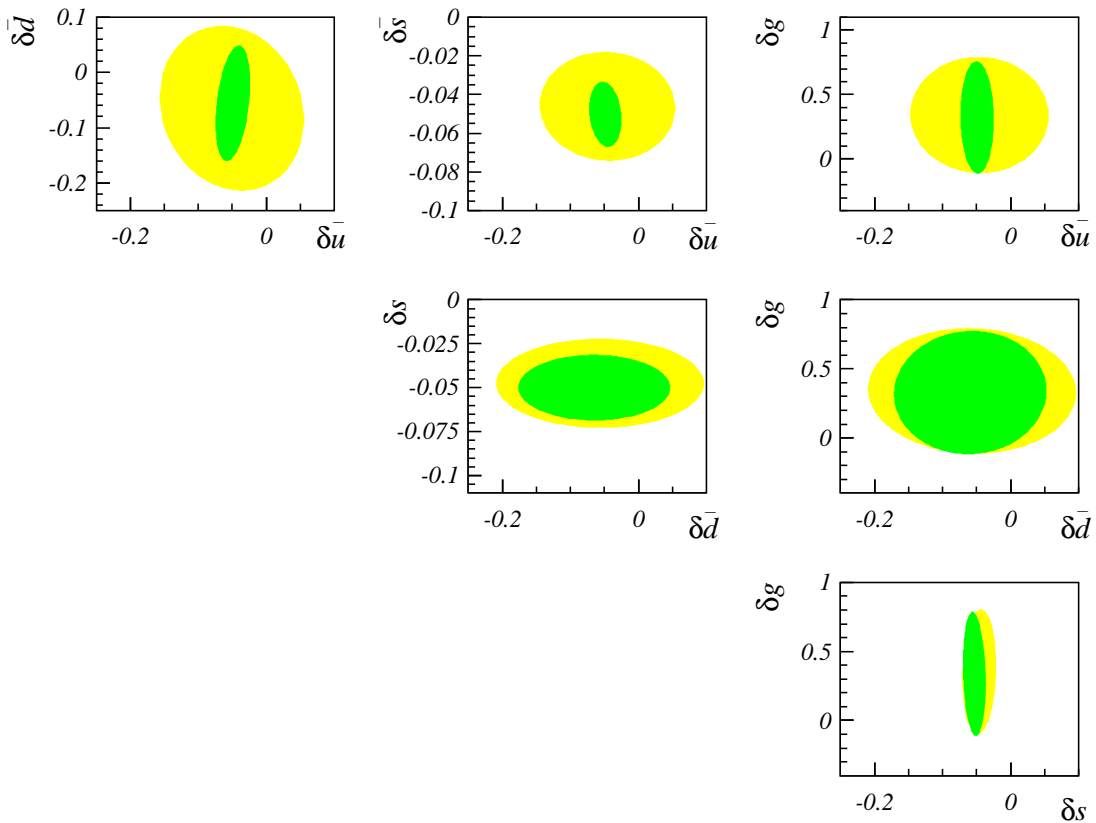}
\caption{Correlations between the flavor polarizations within a  5\% increase 
in $\chi^2$ in both the standard and the improved fits.}
\label{fig:eli}
\end{figure}

As expected, the most noticeable effect is in $\overline{u}$ polarization 
due to the maximal sensitivity of the semi-inclusive asymmetries on proton
targets to this distribution, as discussed in \cite{deFlorian:2000bm}. 
For the $\overline{d}$ and $\overline{s}$ distributions, the impact of the 
future measurements is comparatively suppressed by the weakness of the 
electric charge factor associated with these flavors, however there is a 
noticeable improvement for the $\overline{s}$ quark uncertainty near 
the minimum. In the previous analysis this distribution was mainly constrained 
by positivity resulting in flat $\chi^2$ distribution around the minimum but 
here shows a nice parabolic profile. Notice that both in the 
analysis of reference \cite{deFlorian:2000bm} and in the present one, we are 
forced to assume $\Delta \overline{s}=\Delta s$ since there is not enough 
data yet to discern alternatives. This assumption implies a strong constraint 
on $\delta \overline{s}$ even though the measured asymmetries are less 
sensitive to this distribution than to $\delta \overline{u}$ and 
$\delta \overline{d}$. On the contrary, the relation between 
$\delta \overline{u}$ and $\delta {u}$ and the same for $\delta \overline{d}$ 
and $\delta {d}$ comes from the fit. The impact on the gluon 
distribution is not significant and mainly indirect, coming as found in 
reference \cite{deFlorian:2000bm} through the constraints on the sea quark 
distributions, which are now better defined. It is worth mentioning that the 
impact of the kaon data is very mild, being mostly the pion asymmetries 
responsible for the changes.   

As discussed in references \cite{deFlorian:2000bm,martin:2003}, in modern 
extractions of parton densities it is customary to consider alternative sets 
of parton densities within an increase between 2\% and 5\% in $\chi^2$, as a 
conservative estimate for the range of uncertainty of the global fit. In order 
to estimate the corresponding uncertainty range in the computation of a given 
observable, it is customary to take it as the range of variation of the 
observable within the alternative sets. This is precisely what we show in 
Figure \ref{fig:new} where we plot  the uncertainty 
bands of $A_p^{\pi +}$ and  $A_p^{\pi -}$ corresponding to 
$\Delta \chi^2 = 5 \%$ in the improved fits as the area between the dashed 
lines.
For comparison we include in the plots the values for these pion asymmetries 
at the kinematics of the forthcoming Jefferson Lab experiment, computed with 
the set of \cite{deFlorian:2005mw} and which where included as ``data'' in the
``improved fit'', together with the expected error bars. In this way, we can 
see not only the consistency of the results but also the appropriateness of 
the choice of $\Delta \chi^2 = 5 \%$
Similar uncertainty bands obtained for $\Delta \chi^2 = 5 \%$ but in the 
``standard fit'' were found to be twice as large in \cite{deFlorian:2005mw}.

As it was pointed out in the introduction, an important feature to keep in 
mind regarding the range of variation of the polarization of the different 
flavors is that they are correlated. For example, they cannot be expected 
to hold simultaneously; in consequence, for a given  allowed range in 
$\chi^2$, two or 
more flavors may cannot take their respective maximum departures from the 
best fit value together. For this reason it is worth while to study such 
correlations between the different flavors, and how these correlations 
change with the inclusion of additional data. 
This can be done systematically generalizing Equation (1) for more 
than one flavor polarization, with independent Lagrange multipliers, 
scanning the profile of the $\chi^2$ function in the range of variation of 
them.

In Figure \ref{fig:eli} we show the allowed range of polarization within a 5\%
increase in $\chi^2$ in the $\delta \overline{u}-\delta \overline{d}$, 
$\delta \overline{u}-\delta \overline{s}$, $\delta \overline{u}-\delta g$, 
$\delta \overline{d}-\delta \overline{s}$, $\delta \overline{d}-\delta g$, and 
$\delta \overline{s}-\delta g$ planes. In order to simplify the plots we have 
approximated the actual contours by ellipses, the darker ones obtained with 
the ``improved fit'', while lighter being those coming from the 
``standard fit''. Again the most prominent effect is the shrinkage of 
the $\delta \overline{u}$ and $\delta \overline{s}$ uncertainty range. 

In these plots the correlations between the polarization of the different 
flavors are represented by the angles between the axes of the ellipses and the 
coordinated axes. A positive or negative $\pi/4$ difference would imply a 
maximal positive or negative correlation respectively, and that both 
polarizations are weakly constrained. This is the case, for example for
$\delta \overline{u}$ and $\delta \overline{d}$ in the  ``standard fit'', 
situation that is corrected in the improved version. In the remaining cases, 
the axes of the ellipses are almost parallel to those of the coordinates 
suggesting mild correlations between the different pairs of flavors.

There is however, a remaining subtle correlation between  
$\delta\overline{u}$ and $\delta\overline{d}$ and between 
$\delta \overline{u}$ and $\delta\overline{s}$, that the enlarged set of 
asymmetries included in the fit is still not able to remove. 
In the first case the residual correlation is positive, while in the latter 
it is negative. 
The gluon polarization also seams to have negligible correlation with that 
of the anti-quarks, with only a very slight positive tendency with 
$\delta\overline{s}$ in the standard fit, which is removed in the improved 
one.

\setlength{\unitlength}{1.mm}
\begin{figure}[hbt]
\includegraphics[width=9cm]{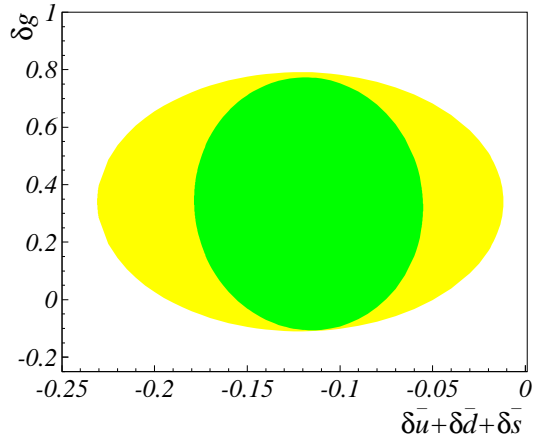}
\caption{Correlation between $\delta \Sigma_{\overline{q}}$ and $\delta g$.}
\label{fig:seatot}
\end{figure}

Another interesting correlation to investigate is the one between the total
polarization carried by anti-quarks and that of the gluons. This in practice
amounts to associate a Lagrange multiplier in Equation (1) to the sum over 
the anti-quark species $\delta \Sigma_{\overline{q}}=(\delta \overline{u}+
\delta \overline{d}+\delta \overline{s})$ and another to the gluon 
polarization.
In Figure \ref{fig:seatot} we see that there is no significant correlation 
between $\delta \Sigma_{\overline{q}}$ and $\delta g$. Notice also how 
significantly the asymmetries expected to be measured by E04-113 will help 
to constrain the anti-quark polarization.

\section{Conclusions}
We have analyzed the potential impact of forthcoming  semi-inclusive 
polarized deep inelastic scattering proton measurements in the determination of 
sea quark polarization in the nucleon by means of a next to leading order  
global QCD analysis. We find that the inclusion of this data
will effectively contribute to constrain the sea quark
polarization in the proton. The most significant improvement is found
in the up anti-quark distribution, and with an also noticeable effect for
strange anti-quarks. For down anti-quarks, the new data will have a smaller 
but nonnegligible effect.  
Regarding the correlations, we found that the forthcoming data will reduce
the apparent correlation found between $\delta \overline{u}$ and $\delta 
\overline{d}$ in the standard fit of reference \cite{deFlorian:2005mw} 
leading to a picture where the sea quark densities and their uncertainties 
can be determine independently. 
\section{Acknowledgements}

We warmly acknowledge Daniel de Florian for comments and suggestions.
R.S. is grateful to Jefferson LAB for the hospitality during his visit
where this analysis was completed. This work was partially supported by 
CONICET, Fundaci\'on Antorchas, UBACYT and ANPCyT, Argentina and the 
US National Science Foundation grant PHY 03-54871.

\end{document}